# Elastic higher-order topological insulators with quantization of the quadrupole moments

Zhen Wang,[1] Qi Wei,[1] Heng-Yi Xu,[1,*] and Da-Jian Wu[1,*]

[1]*Jiangsu Key Laboratory of Opto-Electronic Technology, School of Physics and Technology, Nanjing Normal University, Nanjing 210023, China*

*wudajian@njnu.edu.cn; hengyi.xu@njnu.edu.cn

We demonstrate that HOTIs with the quantization of the quadrupole moments can be realized in the two-dimensional elastic phononic crystals (PnCs). Both one-dimensional (1D) topological edge states and zero-dimensional (0D) topological corner states are visualized and can be transformed each other by tuning the crystalline symmetry in a hierarchical structure. The systematic band structure calculations indicate that elastic wave energy in the hierarchical structure can be localized with remarkable robustness, which is very promising for new generations of integrated solid-state phononic circuits with a great versatility.

*Introduction*.—In recent years, the topological insulators (TIs), which feature the backscattering-immune edge states, have been observed in a variety of quantum Hall families of both fermionic and bosonic systems [1-21] (e.g. quantum Hall effects(QHE), quantum spin Hall effects(QSHE) and the quantum valley Hall effects(QVHE)) and provide building blocks of various topological devices [22-27]. While the higher-order topological insulators (HOTIs), constituting a new family of topological phases, have attracted enormous attention due to their potential as a new type of information carriers with the quantized multipole polarization and enriched our knowledge of nontrivial topological insulating phases. According to the bulk-boundary correspondence, a $d$D TI with $(d-1)$D, $(d-2)$D, ..., $(d-n-1)$D gapped boundary states and $(d-n)$D gapless boundary states is defined as the so-called nth-order TI. Recent theoretical studies have shown that it is possible to realize HOTIs beyond the traditional bulk-boundary correspondence (the first-order TIs) [28,29]. The Higher-order topological corner states can either stem from the quantization of dipole moments, or the quantization of the quadrupole moments [30]. The HOTIs with the quantization of the dipole moments in photonic crystals was proposed by Xie *et al*. [31], and Fan *et al*. [32] implemented the elastic HOTIs with the quantization of the qradrupolee moments in a two-dimensional (2D) breathing graphene lattice. Furthermore, Zhang *et al*. [33] observed experimentally the HOTIs with the quantization of the dipole moments in tunable two-dimensional sonic crystals. Elastic phonons in solids are nowadays advantageous to the high signal-to-noise ratio information processing compared to the fluid/airborne ones. Based on the 2D extension of the Su-Schrieffer-Heeger (SSH) lattice [34–37], we here demonstrate that the quantization of quadrupole moments in the elastic systems can also give rise to HOTIs. The Implementation of the HOTIs with the quantization of quadrupole moments in a Lamb-wave system has two major steps: (I) By introducing a Dirac mass $m_{2D}$, a complete a band gap is opened up in vicinity of the original quadruple degeneracy Dirac point accompanied with a bulk topological transition. And topological edge states emerge at the boundaries between two PnCs with opposite signs of the Dirac mass $m_{2D}$. (II) By breaking the glide symmetry at the domain boundary, edge topological transition is discovered and topological corner states (Jackiw–Rebbi soliton states [38]) show up in the band gap.

The macroscopic controllability of PnCs makes it a versatile tool for exploring acoustic analogues of quantum topological phases in condensed-matter materials in which topological transitions often demand sophisticated atomic-scale manipulations. On the other hand, the elastic wave propagation in well-structured phononic systems exhibits many exotic properties from negative refraction [39], super focusing [40] to cloaking [41]. It is therefore worthwhile to investigate the HOTIs in Lamb-wave PnCs. In this work, we engineer a hierarchical structure formed by two pieces of PnCs, where its inner and outer PnCs have different topological phases, to visualize both the first-order and the second-order topological insulating phases. Various topological transitions between the corner, edge and bulk states by tuning the crystalline symmetry are studied, and the robustness of Lamb-wave HOTIs with topologically protected corner modes are verified to pave a way to manipulate the propagation of Lamb waves and design novel phononic devices.

This paper is organized as follows. Bulk topological phase transition in Lamb-wave PnCs is introduced in Sec. II. In Sec. III and IV, we study the edge topological phase transition and high-order topological phase by the glide symmetry broken in the edge. In Sec. V, we verify the robustness of these topological corner states against defects. A hierarchical topological insulating phases in PnCs is demonstrated in in Sec. VI. Finally, a summary is given in Sec. VII.

*Bulk topological phase transition*.—Our 2D PnCs consist of a squared lattice with four artificial atoms in a unit cell as depicted in Fig. 1(a), where $l_{1,2}$, $w_{1,2}$ and $h_{1,2}$ are the length, width and height of the plate and prisms, respectively. $\theta$ is rotation angle of a quadrangular prism as shown in Figs. 1(b) and 1(c). Solid stainless steel prisms (blue) are arranged as a squared-lattice structure on the surface of aluminum plate (red). The constitutive parameters of stainless steel and aluminum plate in the calculations are shown in Table I. Full-wave simulations are carried out by a commercial finite-element solver (COMSOL Multiphysics). In the calculations, the free boundary conditions on the upper and lower surfaces of the plate and Floquet periodicity around the unit cells are applied, which ensure validity of 2D approximations. The out-of-plane modes (black lines) characterized by the parabolic dispersion are loosely coupled with in-plane modes (gray lines), which is beyond the scope of this work and is neglected [32]. When the geometric parameters of prisms and plate are identical [$l_1=w_1=a$, $h_1=0.4a$, $l_2=0.35a$, $w_2=0.15a$, $h_2=0.35a$, $a$ (25μm) is the lattice constant], there exists a quadruple degenerate point ($\theta=0°$) at high symmetry point X as shown in Fig. 1(e) with the geometry of the first Brillouin zone being depicted in the inset of Fig. 1(e). A Dirac mass $m_{2D}$ is introduced to tight-binding model (2D SSH model) by rotating the angle $\theta$ of the four meta-atoms such that the degeneracy of the gapless point is lifted and a full phononic band gap emerges as presented in Figs. 1(c) and 1(e). We further consider the cases of the inverted and eversion lattices [indicated by blue prisms in Figs. 1(b) and 1(c)]: with $\theta=\pm45°$, which corresponds to the maximum value of bulk band gap [33] while preserves the mirror symmetry. As $\theta$ is varied gradually from -45° to 45°, the bulk band gap will undergo a topological phase transition as shown in Figs. 1(c)-(e). Moreover, lattice ($\theta<0°$) and eversion lattice ($\theta>0°$) are characterized by two distinct topological phases which are connected by a band-inversion process.

TABLE I. Parameters of aluminum and stainless steel used in numerical simulations.

|  | aluminum | stainless steel |
| --- | --- | --- |
| Density (kg/m$^3$) | 2700 | 7903 |
| Young's modulus (GPa) | 70 | 219 |
| Poisson's ratio | 0.34 | 0.32 |

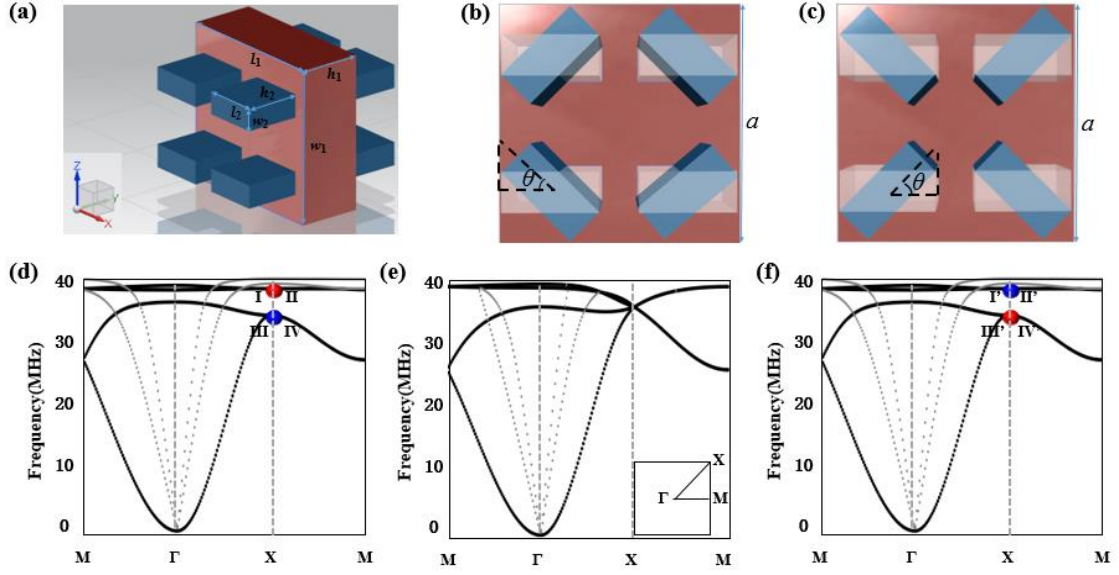

Fig. 1. Crystal structures and bulk band structures. (a) Three-dimensional structure of a unit cell with aluminum plate (red) and stainless steel prisms (blue). The illustration is a Cartesian axis. $l_{1,2}$, $w_{1,2}$ and $h_{1,2}$ are the length, width and height of the plate and prisms, respectively. (b) and (c) Projection of the unit cell in the y direction. The white translucent area represents the normal square lattice ($\theta=0°$). The blue areas are inverted (b) and everted (c) lattices with $\theta=-45°$ and $\theta=45°$, respectively. (d) Band structures for $\theta=-45°$(left), $\theta=0°$(middle), and $\theta=45°$(right), which represent inverted, normal, and everted lattice, respectively. For inverted and everted lattices, the band structures have the same dispersion but different parities (denoted by red and blue symbols at the X point).

To get a deeper physics insight of the above phenomenology, we continue to analyze the elastic field distributions of the unit cell for two cases ($\theta=\pm45°$) as shown in Fig. 2, where the black arrows represent the directions of elastic energy flow. The patterns in Figs. 2 (a) and 2(b) correspond to the modes at the high symmetry point X in Figs. 1(d) and 1(f), respectively. By analyzing the patterns on the surface of the four scatters of unit cell I in Fig. 2(a), we find that the stainless steel prisms evolve via two different manners: the first one vibrates in a radially breathing manner between prisms A and D, and the second one characterizes a peer chiral elastic energy flow rotating in the xz-plane in the time-domain between prisms B and C. In addition, it is evident that every two modes with the same frequency show opposite chirality on the surface of the aluminum plate, so that the chiral elastic energy flows can be viewed as pseudospins [42-44]. By utilizing these analogues, researchers have achieved electromagnetic [45-47] and acoustic transmission line [48-51] with robust, valley-dependent transport of wave energy.

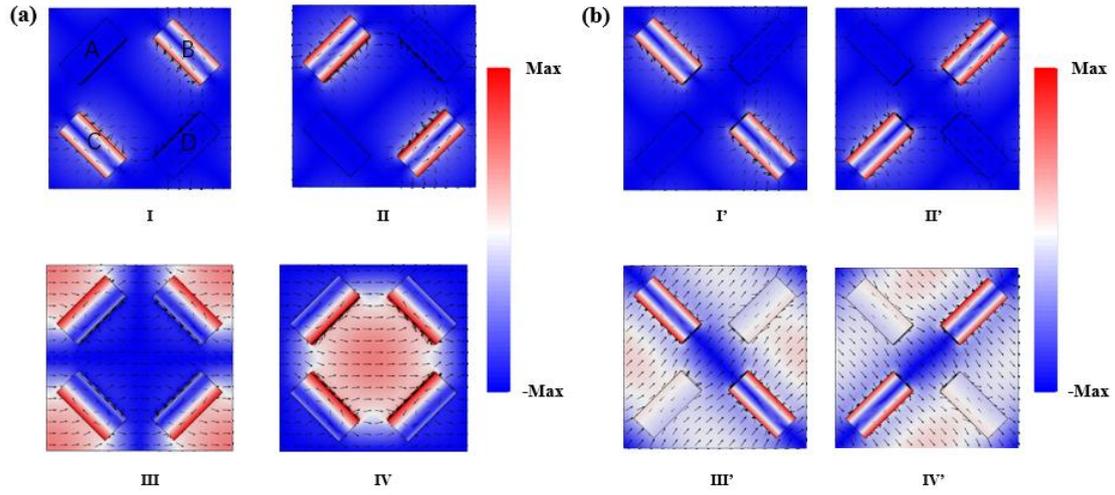

Fig. 2. The elastic field distributions in two different unit cells. (a) Elastic field distribution inside a unit cell when $\theta=-45°$, upper and low panels correspond to the red and blue symbols in Fig. 1d, respectively. (b) Elastic field distribution inside a unit cell when $\theta=45°$, upper and low panels correspond to the blue and red symbols in Fig. 1d, respectively.

*Edge topological phase transition.*—To demonstrate the existence of topological edge states, we consider a composite structure of PnC where a topologically nontrivial PnC is jointed by a topologically trivial PnC. The simulated projected band structure is shown in Fig. 3(a). In the simulations, the absorption boundary conditions are applied on the surfaces parallel to the interface of two PnCs to exclude extra edge states, while Floquet periodic boundary conditions are used for the boundaries perpendicular to the interface. All the other parameters remain the same as those in Fig. 1. The red and blue triangle marks represent the equivalent elastic pseudospin up and pseudospin down, respectively. Their corresponding one-dimensional local modes and the elastic energy flow are displayed in Figure 3(c) as well as the inset (black arrows). It is evident that the elastic energy flows of pseudospin states evolve in the opposite directions albeit their phase distributions are the same. Furthermore, the 1D localized states for the nontrivial cases emerge and decay rapidly away from the interface.

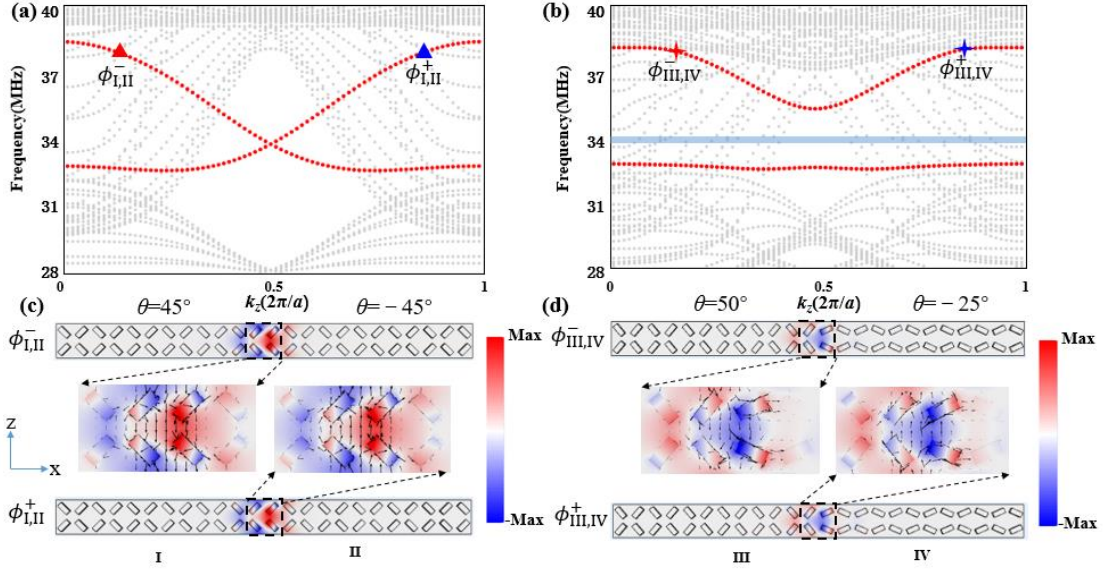

Fig. 3. Edge pseudospins and gapped helical edge states. (a) The dispersions of the bulk (grey curves) and edge (red curves) states for supercells with boundaries along the $z$ directions between the two PnCs with $\theta=\pm 45°$, respectively. (b) The dispersions for supercells with boundaries along the $z$ directions between the two PnCs with $\theta = -25°$ and $50°$, respectively. (c) elastic displacement profiles for the edge states marked by the red and blue triangle marks in a (a). (d) elastic displacement profiles for the edge states marked by the red and blue cross marks in a (b). Poynting vectors (black arrows) shown in the inset of (c) and (d).

The pseudo-Kramers double degeneracy is preserved under the glide operation $\left[G_x \coloneqq (x,z) \to \left(\frac{a}{2}+x, \frac{a}{2}-z\right) \text{ and } G_z \coloneqq (x,z) \to \left(\frac{a}{2}-x, \frac{a}{2}+z\right)\right]$ near the interface (black dotted box) in Figs. 3(a) and 3(c). When the glide symmetries are lifted, the elastic edge states are no longer gapless and a bandgap is opened up at the $k_z=\pi/a$ point as shown in Fig. 3(b). Such gapped edge states resembling the edge states in QSHIs can be well described by the 1D massive Dirac equations [53,54]. For instance, an omnidirectional band gap at the original degenerate point is formed along the edges between the two PnCs with $\theta$=-25° and 50°. These one-dimensional local modes corresponding to the cross marks, as well as the elastic energy flow directions are shown in Fig. 3(d). Analogous to the case in (c), the energy flow directions are reversed between $\phi^-_{III,IV}$ and $\phi^+_{III,IV}$ even though the phase distributions of the two modes remain unchanged. The dispersions of the lower edge states have much smaller group velocity and hence much longer propagation time over the same distance. Therefore, the lower edge states are suppressed and the dispersion resolution is reduced considerably. However, because the glide symmetries are broken on the edges, the edge states of pseudo-spin up and down are coupled and mixed each other which can be envisioned as a remnant effect of the pseudo-spin-momentum locking for helical edge states in QSHIs [52].

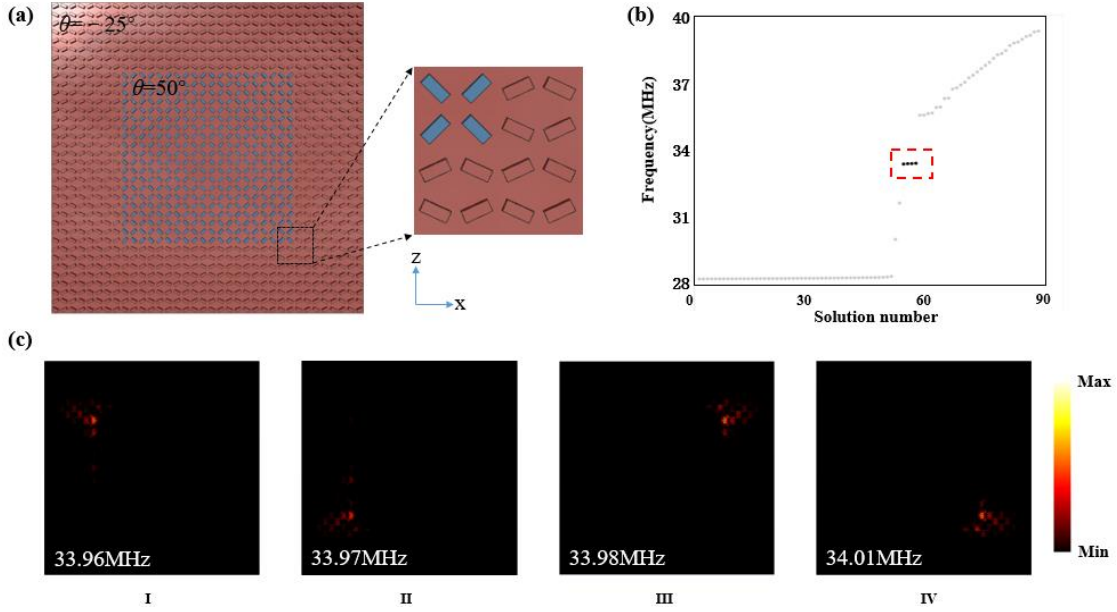

Fig. 4. The visualization of corner states in a metastructure. (a) The CAD drawing of a metastructure with outer PnCs (red, $\theta$=-25°) and inner PnCs (blue, $\theta$=50°). The enlarged corner structure is presented in the inset. (b) Eigenmodes calculation of the metastructure with the same parameters in Fig. 1. Corner states are represented by blue dots in a red dotted box. (c) The elastic energy distribution of the four corner states in (b).

*Corner states in in a hierarchy of dimensions*.—We now extend the discussion from the previous hybrid structure to a squared meta-structure as shown in Fig. 4(a). Here, the bulk-edge-corner correspondence is manifested in a hierarchy of dimensions: the bulk topology leads to the edge states, while the edge topology leads to the corner states, as shown in Figs. 3(b) and 4(a) representing the smokinggun feature of SOTIs. The corner states in a box-shaped geometry is formed by the two PnCs with $\theta$=-25° and $\theta$= 50° and arise due to the breaking of glide symmetry in $C_4$-symmetric topological crystalline insulators [55]. The frequencies of eigenmodes with absorbing boundary condition in the x and z directions to avoid the formation of standing waves are shown in Fig. 4(b). The four corner states emerge in the edge bandgap labeled by the blue points in the red dotted box with the frequencies 33.96MHz, 33.97MHz, 33.98MHz, 34.01MHz, respectively. It is worth mentioning that there is a frequency difference between every two corner states in the elastic-wave system, unlike the degeneracy occurring in the photonic system [31]. The elastic field (polarized in the y direction) distribution shown in Fig. 4(c) clearly demonstrates that these four states are strongly localized at four corners of the box-shape boundary in the meta-structure, indicating the subwavelength character of the corner states.

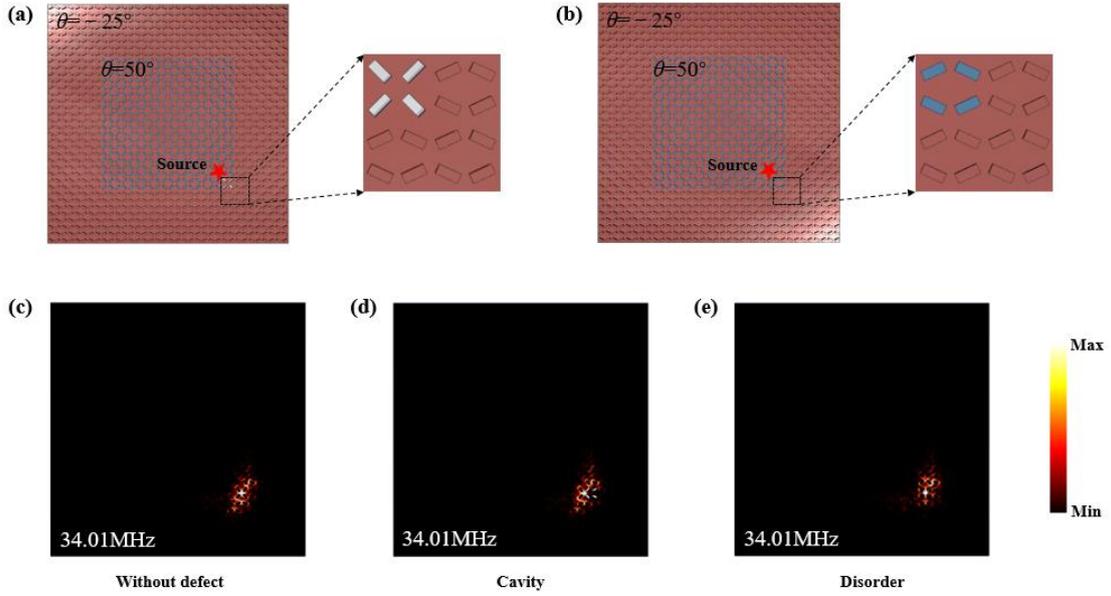

Fig. 5. The study on the robustness of the corner states. (a) and (b) The CAD drawing of a metastructure with cavity and disorder on the bottom-right corner. In the inset of (a) and (b), the enlarged defect structures are presented. The red stars are the positions where the exciting sources are applied on. (c), (d) and (e) The elastic energy distribution of the square metastructure without defect, with cavity and disorder, respectively.

*Robustness of the topological corner states.*—In this section, we perform a systematic study on the robustness of the corner states against defects based on simulations. Here we consider the excitation under a source (an initial displacement along the y direction) near to the bottom-right corner as shown by the red stars in Figs 5(a) and 5(b), for a sample with 22×22 unit cells (the same as that used in Fig. 4(a)) to reduce the finite size effect. The pristine structure has a box geometry with θ=-25° and θ=50° and the corresponding elastic energy distribution are shown in Fig. 4(a) and Fig. 5(c), respectively. The two types of defects are introduced on the corner including the cavities and disorder. Cavities are modeled by removing four scatters in the bottom-right corner of the blue area as shown in the inset of Fig. 5(a), and the corresponding elastic energy distribution is shown in Fig. 5(d). The case of disorder is shown in the inset of Fig. (b) where $\theta$ is changed from 50° to -25° in the bottom-right corner of the blue region with the corresponding elastic energy distribution in Fig. 5(e). It is found that the elastic energy near the topological corner states is well localized and not scattered to the one-dimensional boundary or the interior of the PnCs as the conventional edge/corner states [56]. And the robustness of the corner states against various defects is fully confirmed in our calculations.

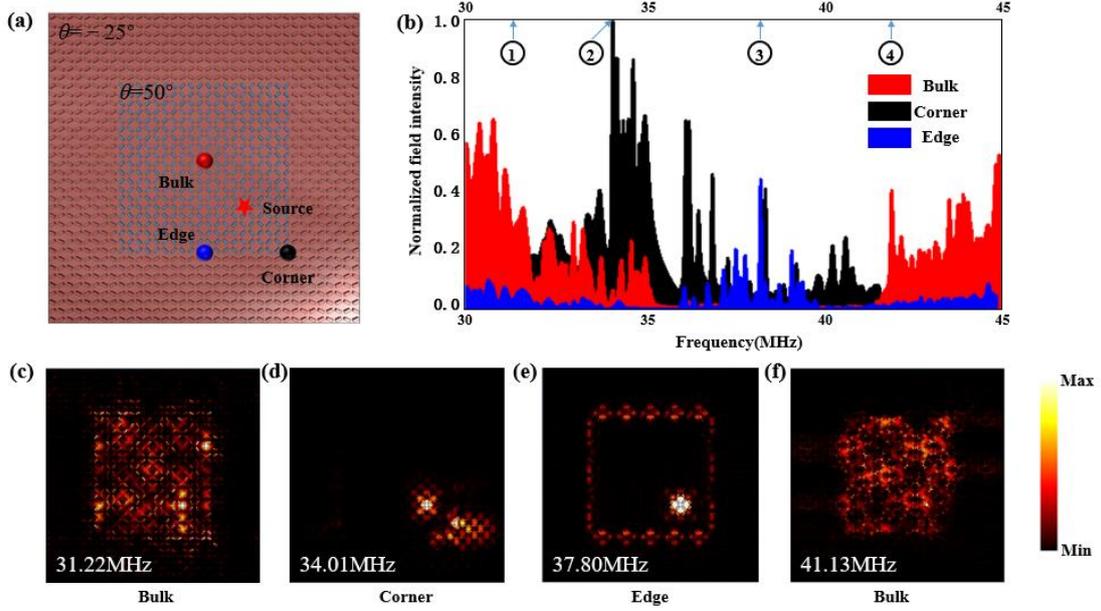

Fig. 6. Hierarchical structure of topological insulating phases. (a) The hierarchical structure with three sampling positions of bulk (red), corner (black) and edge (blue). The red star represents the exciting sources. (b) Local energy intensity calculated at bulk, edge, and corner points as depicted in (a). The bulk, corner, and edge eigenstates from the numerical calculation are represented by red, black, and blue waterfall maps, respectively. We provide the visualization of (c) bulk state at 31.22 MHz, (c) corner state at 34.01 MHz, (d) edge state at 37.80 MHz, and (e) bulk state at 41.13 MHz with the excitation source placed at the bottom-right corner of the meta-structure.

*Hierarchical topological insulating phases.*—To further verify the coexistence of 1D topological edge states and 0 D topological corner states, we excite the eigenstates from 30MHz to 45MHz as shown in Fig. 6. The local energy intensity for bulk, edge, and corner points (denoted by respective red, blue, and black dots in Fig. 6(a)) are obtained from the averaged values of the energy distributions at each point. The red star represents the exciting sources. The spatial distributions of the states corresponding to the respective frequency points 1-4 exhibit a bulk-corner-edge-bulk evolution as we sweep the excitation frequencies. In Fig. 6(d), the energy strength around the 0D corners is rather strong, while the energy strength on the 1D boundaries as well as 2D bulk of the structure (except from the source) are relatively weak, which is consistent with the characteristics of the SOTI phases. Similarly, the energy strength of the 2D bulk of the meta-structure in Fig. 6(e) is significantly suppressed, representing the gapped 2D bulk states. In contrast, a strong field strength at the 1D boundaries signaling the presence of the gapped 1D edge states is apparently visible. Since the 1D boundary and the 0D corner states coexist at the point 3, the two resultant peaks in the elastic energy nearly overlap as shown in Fig. 6(b). These features conform with the traditional bulk-boundary correspondence and the definition of the first-order topological insulators.

*Conclusions.*—In summary, we here demonstrate a 2D SOTI in elastic PnCs and visualize both 1D topological edge states and 0D subwavelength corner states by modulating the crystalline symmetry in a square lattice. Meanwhile, our realization is based on PnCs with elastic modulus varying with periodicity, which have wider bandgap than that composed of a single material. The hierarchical structure of topological insulating phases is observed in a topological nontrivial configuration [57]. Moreover, the coexistence of different dimensional topological boundary states can serve as a basis for designing topological switch circuits between crystalline insulators and HOTIs [1]. Our work exhibits a good ability to control the elastic wave propagation in an unprecedented way and provides a platform to design the new type of elastic topological devices which can topologically transform the elastic wave energy among bulk, edge, and corner modes by utilizing advanced microfabrication technology [58].

This work was supported by the National Natural Science Foundation of China (Nos. 11674175, 11874222, and 11704193) and "333" Project of Jiangsu Province (No. BRA2017451).